\documentclass{article}
\usepackage{graphicx} 
\usepackage[a4paper, total={4in, 8in}]{geometry}
\usepackage{url}
\usepackage{cite}

\title{Custom Data Augmentation for low resource ASR using Bark and Retrieval-Based Voice Conversion  \\\line(1,0){\textwidth}}
\begin{document}
\author{
  Anand Kamble\\
  \texttt{\scriptsize Department of Scientific Computing}\\
  \texttt{\scriptsize Florida State University, USA}\\
  \texttt{\scriptsize amk23j@fsu.edu}
  \and
  Aniket Tathe\thanks{Corresponding author. Email: \scriptsize anikettathe.08@gmail.com}\\
    \texttt{\scriptsize Department of  Mechanical Engineering}\\
  \texttt{\scriptsize MES College of Engineering, Pune, India}\\
   \texttt{\scriptsize  anikettathe.08@gmail.com}
  \and
  Suyash Kumbharkar\\
    \texttt{\scriptsize Department of Electrical Engineering }\\
    \texttt{\scriptsize and Information Technology}\\
  \texttt{\scriptsize Technische Hochschule Ingolstad, Germany}\\
  \texttt{\scriptsize suk9387@thi.de}
  \and
  \\
  Atharva Bhandare\\
    \texttt{\scriptsize Department of  Mechanical Engineering}\\
  \texttt{\scriptsize MES College of Engineering, Pune, India}\\
  \texttt{\scriptsize atharvabhandare512@gmail.com}
  \and
  \\
  Anirban C. Mitra\\
    \texttt{\scriptsize Department of  Mechanical Engineering}\\
  \texttt{\scriptsize MES College of Engineering, Pune, India}\\
  \texttt{\scriptsize amitra@mescoepune.org}
}

\date{}

\maketitle
\section{Abstract}
This paper proposes two innovative methodologies to construct customized Common Voice datasets for low-resource languages like Hindi. The first methodology leverages Bark, a transformer-based text-to-audio model developed by Suno, and incorporates Meta's enCodec and a pre-trained HuBert model to enhance Bark's performance. The second methodology employs Retrieval-Based Voice Conversion (RVC) and uses the Ozen toolkit for data preparation. Both methodologies contribute to the advancement of ASR technology and offer valuable insights into addressing the challenges of constructing customized Common Voice datasets for under-resourced languages. Furthermore, they provide a pathway to achieving high-quality, personalized voice generation for a range of applications.

\section{Keywords}
Bark, Retrieval-Based Voice Conversion, Custom Voice Cloning, Text-To-Speech, Data Augmentation.

\section{Introduction}

The Common Voice corpus\cite{ardila2020common} is a comprehensive and highly diverse collection of transcribed speech, boasting an extensive repository of 19,160 validated hours spanning 114 different languages. Each dataset entry comprises a unique MP3 audio file coupled with its corresponding text transcript. Moreover, this resource goes a step further by including valuable demographic metadata, such as age, gender, and accent, in a substantial portion of the 28,751 recorded hours. This additional metadata can prove invaluable for enhancing the accuracy of speech recognition systems.

Despite its vast size and inclusivity, the Common Voice corpus still presents challenges when dealing with low-resource automatic speech recognition (ASR) languages like Hindi. In the latest release, Common Voice 15.0, English language data encompasses approximately 3,347 hours of audio, of which 2,532 hours have been meticulously validated by a community of 88,904 contributors. In stark contrast, the available Hindi dataset comprises a mere 20 hours of recorded speech, with just 14 hours validated and contributions from 396 unique voices, significantly lagging behind in terms of volume and diversity. This stark contrast underscores the need for further attention and development in low-resource language ASR and TTS research.

Significant research efforts are currently dedicated to advancing the realm of Text-to-Speech (TTS) technology. Text-to-speech (TTS) is the task of generating natural-sounding speech given text input. Text-to-speech (TTS) models can be used in any speech-enabled application that requires converting text to speech imitating the human voice. Some of the models are Tortoise TTS\cite{betker2023better}, Bark\cite{suno-bark}, Tachotron\cite{wang2017tacotron}, Tachotron 2\cite{8461368}, FastSpeech2\cite{ren2022fastspeech}. Bark is not a conventional text-to-speech model but instead a fully generative text-to-audio model. Bark was developed by Suno which is a transformer-powered text-to-audio system capable of producing exceptionally lifelike speech in multiple languages. Additionally, Bark has the capacity to generate various types of audio content, ranging from music and ambient sounds to basic sound effects. Moreover, the model can create nonverbal expressions such as laughter, sighs, and crying.
The methodology presented in this paper focuses on generating a custom corpus similar to Common Voice using small audio clips using Bark, Meta’s Encodec, and HuBert which can be used for personalized tasks and even for custom ASR when dealing with low-resource ASR language like Hindi.

\section{Methodology 1}
\subsection{Preprocessing}

First, a video from YouTube was scraped using pytube library. Then the video is converted into audio using moviepy (A Python library that uses FFmpeg for video editing and manipulation). If the audio contains noise, using noisereduce library the noise can be removed from the audio. Additionally 
Spleeter (open-source audio source separation library developed by Deezer) allows you to split audio signals into constituent audio tracks, such as vocals, accompaniment, drums, etc. This is also called stem separation. Spleeter provides several pre-trained models with different stems (2stems,4stems and 5stems). According to the audio input suitable Spleeter\cite{Spleeter} with suitable stems can be selected for splitting purposes and vocals extracted from it can be used ahead. These methods are optional and can be used as a part of preprocessing if the audio contains noise. Since in our case the audio did not contain significant noise these steps were omitted. The audio is then trimmed into audio clips of lengths 5,10 and 15 sec and saved into three separate folders. These were further used to see how the clip length affect the overall model performance during audio generation. These trimmed audio clips are then converted from MP3 and wav files.

\subsection{Audio Codebook extraction and Semantic token generation.}

Initially, Discrete tokens are generated from the audio codebooks by utilizing Meta's encoded, and fine, coarse prompts are saved. We make use of Meta's EnCodec to carry out the extraction of audio codebooks from the supplied source audio. Consequently, we attain high-resolution audio embeddings that encompass more nuanced audio details compared to conventional spectrogram-based features. These codebooks comprise two tiers of information, fine and coarse, providing the flexibility to balance precision and compression rate according to preference. In the endeavor to generate semantic tokens aligned with the source audio, our methodology involves utilizing the transcript of the audio in conjunction with the original BARK model. It is important to note, however, that this process encounters limitations attributed to the unavailability of the wav2vec model and its associated kmeans utilized in the initial training. To address the constraints inherent in the semantic token generation process, we address this challenge by incorporating a pre-trained HuBert model featuring a linear projection head. This model is trained to produce tokens within the same embedding space as the wav2vec model, which is not directly accessible\cite{Schumacher2023}.



\section{Methodology 2}
\subsection{Preprocessing using ozen toolkit}

A 1-hour video was scraped from YouTube and converted into WAV format. Subsequently, the Ozen toolkit\cite{Ozen-toolkit} was employed to extract speech, transcribe using Whisper, and save the results in the LJ format. The audio files are stored in the 'wavs' folder, while the transcribed Hindi texts are placed in the 'train' and 'valid' text files. It's important to note that Whisper defaults to English transcription. To transcribe the audio into Hindi, modifications, such as changing the "task": "transcribe" and "language": "hi" were made in the utils.py. for Hindi language. Additionally, the Ozen toolkit incorporates pyannote\cite{bredin2021endtoend} for speaker diarization, enhancing its utility in speech and audio processing applications.

\subsection{RVC model training and inference}

The RVC-Project/Retrieval-based-Voice-Conversion-WebUI is an open-source project that allows users to convert voice data into different voices using a retrieval-based voice conversion approach. The project can be found on GitHub\cite{RVC} and is available for free under the MIT software license. It is recommended that audio recordings be a minimum of 10 minutes long for model training. If the dataset contains noise, it can be removed through Vocals/Accompaniment Separation and Reverbation using the 'HP2-all-vocals' model. The target sample rate was set to 32,000. The base v2 pre-trained model was used for training (f0G32K and f032K) with a batch size 40 for 200 epochs on the NVIDIA A5000 GPU. After training for 200 epochs at 40 batch size, the KL Divergence Loss came down to 0.6777.

\begin{figure*}[ht]
    \centering
    \includegraphics[width=0.5\textwidth]{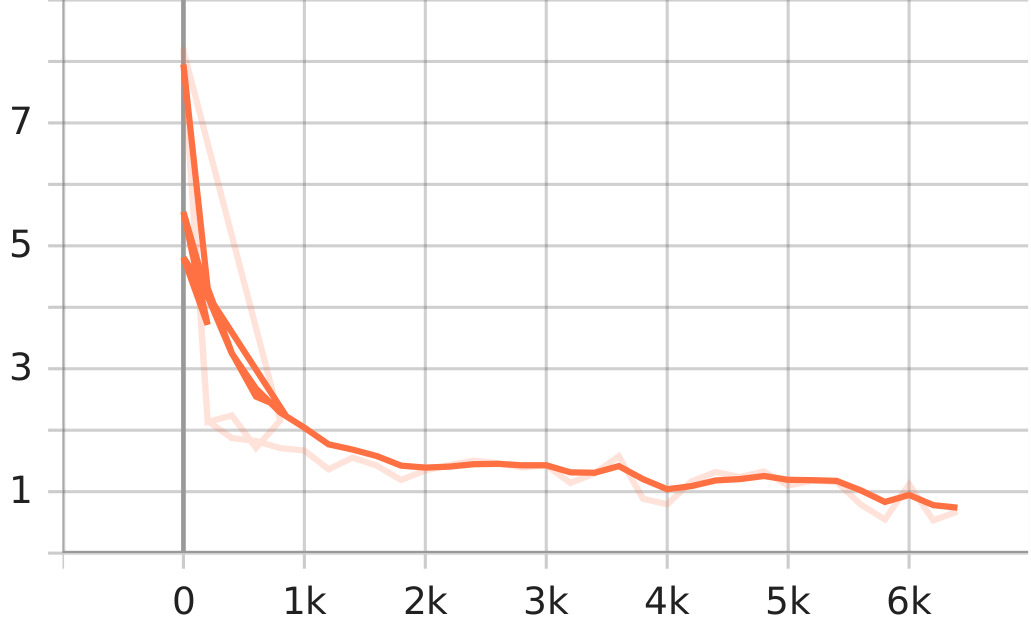} 
    \caption{KL Divergence Loss (loss\_kl)}
    \label{fig:enter-label}
\end{figure*}

Upon completing the training process, the model was used for inference to replicate the trained voice. Common voice audios were then used as inputs with the trained inferencing custom voice and index path, resulting in the creation of a custom common voice dataset with the desired custom voice. Parameters such as volume envelope scaling (0.25), filter radius (3), and search feature ratios (0.75) and 0.33 for the Protection of voiceless consonants and breath sounds. These parameters may vary depending on the audio used for training, and it is recommended to test various settings to find the optimum ones for the user. The generated custom common voice 11.0  dataset with the custom voice can be viewed at “Aniket-Tathe-08/Custom\_common\_voice\_dataset\_using\_RVC” on the hugging face.

\begin{figure*}[ht]
    \centering
    \includegraphics[width=1\textwidth]{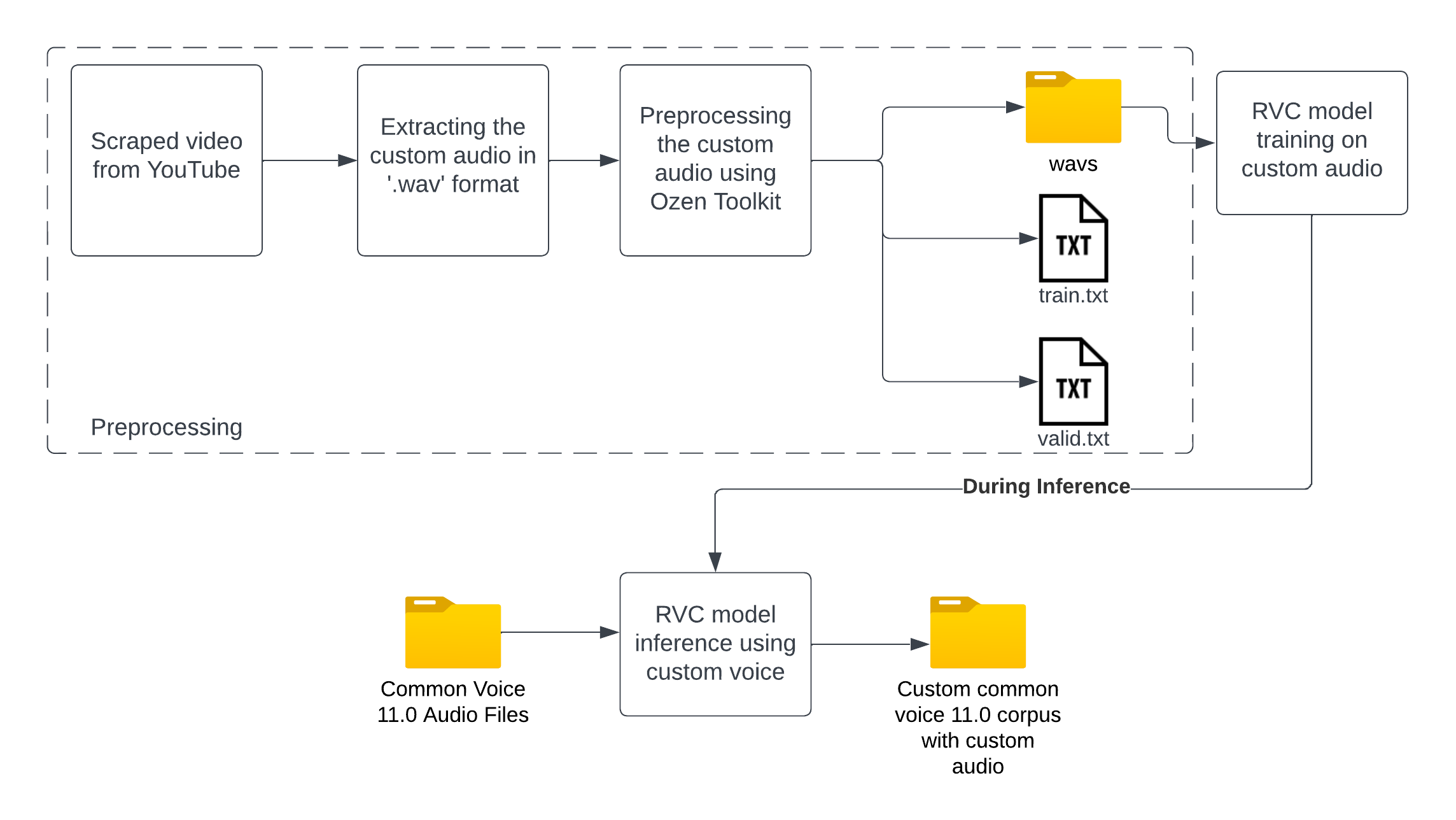} 
    \caption{Model training and Inference}
    \label{fig:enter-label}
\end{figure*}
\newpage
\section{Results and Discussion}
\subsection{Bark}
The resultant tokens are subsequently stored in assets, facilitating their utilization for audio generation by processing the corresponding .npz file. A custom Common Voice dataset can be curated by inputting text and generating the associated audio. Notably, optimal outcomes were achieved when employing a text temp set at 0.85 and waveform temp at 0.7 (these were found using experimentation), wherein the model exhibited commendable performance, occasionally closely resembling the original voice. Furthermore, 10-second clips consistently outperformed their 5-second and 15-second counterparts. It is imperative to acknowledge occasional instances of suboptimal outputs, including instances of undesirable audio generation. Enhancing the model's performance is feasible through rigorous training and fine-tuning of the semantic, fine, and coarse models on the available training data\cite{Serp-ai}.
\subsection{RVC}
When comparing both methodologies, RVC performs exceptionally well, producing an output very close to the custom voice with minimal noise, in contrast to Bark. This makes it a valuable method for generating a custom common voice dataset, suitable for use as a data augmentation technique in custom ASR and low-resource ASR projects. Similar to Common Voice, additional data can also be augmented using OpenSLR and LibriSpeech\cite{7178964}

\nocite{9725035}
\nocite{9915932}
\nocite{9414403}
\nocite{9413889}
\nocite{9413466}
\nocite{8682168}
\nocite{Luong2019TrainingMN}

\bibliographystyle{plain}
\bibliography{references}

\end{document}